\documentclass[11pt,a4paper]{article}
\usepackage{amsmath,amssymb}
\usepackage{epsfig,graphicx}

\newcommand{\Gerda}{{\sc Gerda}}
\newcommand{\Corrado}{{\sc Corrado}}
\newcommand{\GempiIII}{Ge{\sc mpi~III}}

\begin{document}

\title{\bf Highly Sensitive Gamma-Spectrometers of G{\Large ERDA} for Material Screening:
Part 2}

\author{D. Budj\'{a}\v{s}$^a$, W. Hampel$^a$, \underline{M. Heisel}$^a$\footnote{Corresponding author. {\bf e-mail}:
mark.heisel@mpi-hd.mpg.de}, G. Heusser$^a$, M. Keillor$^{a,b}$, M.
Laubenstein$^c$,\\ W. Maneschg$^a$, G. Rugel$^d$, S.
Sch\"onert$^a$, H. Simgen$^a$, H. Strecker$^a$
\\
$^a$ \small{\em Max-Planck-Institut f\"ur Kernphysik,
Saupfercheckweg 1, 69117 Heidelberg, Germany}
\\
$^b$ \small{\em now at: Pacific Northwest National Laboratory, 902
Battelle Boulevard, Richland, WA, USA}
\\
$^c$ \small{\em Laboratori Nazionali del Gran Sasso, S.S.17/bis,
km 18+910, 67010 Assergi (AQ), Italy}
\\
$^d$ \small{\em Technische Universit\"at M\"unchen (E15),
James-Franck-Strasse 1, 85747 Garching, Germany}
\\
}
\date{September 20, 2007}
\maketitle

\begin{abstract}
The previous article about material screening for {\Gerda} points
out the importance of strict material screening and selection for
radioimpurities as a key to meet the aspired background levels of
the {\Gerda} experiment. This is directly done using low-level
gamma-spectroscopy. In order to provide sufficient selective power
in the mBq/kg range and below, the employed gamma-spectrometers
themselves have to meet strict material requirements, and make use
of an elaborate shielding system.

This article gives an account of the setup of two such
spectrometers. {\Corrado} is located in a depth of 15~m~w.e. at
the MPI-K in Heidelberg (Germany), {\GempiIII} is situated at the
Gran-Sasso underground laboratory at 3500~m~w.e. (Italy). The
latter one aims at detecting sample activities of the order
$\sim$\nolinebreak10~$\mu$Bq/kg, which is the current
state-of-the-art level. The applied techniques to meet the
respective needs are discussed and demonstrated by experimental
results.
\end{abstract}

\section{Introduction}

The previous article \cite{part1} points out the relevance of
various material screening procedures employed for the material
selection in scope of the {\Gerda} experiment \cite{gerda}. The
most direct technique for the detection of gamma-active
radioimpurities in the construction and shielding materials is the
low-level gamma-spectroscopy. Like {\Gerda} it is based upon the
operation of high purity germanium detectors (HPGe)\cite{hult},
and hence is sensitive to the same sources of background, out of
which gamma-active isotopes with Q-values above $Q_{\beta\beta}$
(2039~keV for $^{76}$Ge) are of special concern.

In order to meet the aspired background index of
$<$10$^{-2}$cts/(keV$\cdot$kg$\cdot$y) at $Q_{\beta\beta}$ in
Phase~I, and $<$10$^{-3}$cts/(keV$\cdot$kg$\cdot$y) in Phase~II,
sensitivities in material screening ranging from mBq/kg down to
some 10~$\mu$Bq/kg are required. The latter is the current
state-of-the-art level accomplished by the Ge{\sc
mpi}-spectrometer \cite{gempi1, gempi2} located in the Low Level
Research Facility (LLRF) in the Gran Sasso underground laboratory
at 3500~m~w.e.

Apart from reaching the highest possible sensitivities of the
$\gamma$-spectrometers, it is necessary to provide sufficient
screening capacity to cope with the amount of samples to be
screened. Since the measurement time for best sensitivity with
Ge{\sc mpi} can reach up to several months for one sample, we seek
to build new $\gamma$-spectrometers in different depths and
sensitivities.

The \Corrado-spectrometer has been set up in the Low Level
Laboratory (LLL) at 15~m~w.e. at the MPI-K Heidelberg (Germany) as
a follow-up to {\sc Dario} \cite{heusser1, heusser2}. It can reach
a sensitivity of $\sim$\nolinebreak1~mBq/kg corresponding to
10$^{-10}$~g/g for uranium and thorium, and is designed for large
samples, partly for their preselection. {\GempiIII} is successor
of Ge{\sc mpi} at LLRF, aiming for a sensitivity of 10~$\mu$Bq/kg
corresponding to 10$^{-12}$~g/g for uranium and thorium. Both
spectrometers' design has been improved with respect to their
precurser instruments. Their shielding systems suit the demands of
their depth, and are described in the subsequent sections. While
{\Corrado} is operating and first data can be presented,
{\GempiIII} suffers a $^{207}$Bi contamination, which is discussed
in section \ref{contamination}.

\section{Basic Design Criteria}

The central task in the construction of a low-level
$\gamma$-spectrometer is to minimize the lowest detectable
specific activity $A_{min}$ in a certain counting time $t$, as a
measure of it's sensitivity:
\begin{equation}A_{min}=\frac{\sqrt{B}}{M\cdot \epsilon \cdot
t}\label{amin}\end{equation} This equation holds for the simple
case that only one $\gamma$-line of a given isotope is taken into
account for the evaluation, and the background from the line is
negligible. $B$ denotes the background counts in the continuum,
$M$ the sample mass and $\epsilon$ the full energy peak efficiency
associated with the sample.

Minimizing $A_{min}$ means to maximize the potential signal count
rate of the sample, represented by the product $M\cdot\epsilon$.
Therefore a large sample mass $M$ in a geometry preferably
encasing the detector for best efficiency is needed. A large
crystal also leads to a higher efficiency. However, the product
$M\cdot\epsilon$ only increases until self-absorption within the
sample becomes dominant. An optimal sample size can be estimated
from the 1/e thickness for different high density sample material
and the highest energy of interest, in this case the 2615~keV line
of $^{208}$Tl (see \cite{gempi1} for more detail). This results in
the choice of a Marinelli type geometry for the sample chamber
with a volume of about 18~l.

The main room for improvement now remains in the reduction of the
background, which enters (\ref{amin}) as $\sqrt{B}$. Apart from
the continuos background $B$, which can be determined from the
sample spectrum itself, the line background has to be derived from
a separate background spectrum. Its effective rate must be
quantified with regard to the absorption and radon displacement
caused by the sample, which may lead to additional uncertainty
\cite{meinedipl}. Thus it is essential to have low line background
rates and a good control over the chambers radon content. For a
detailed discussion of background sources and suppression
techniques see \cite{heusser3} and \cite{heusser4}.

\section{The C{\normalsize ORRADO}-Spectrometer}
\label{corrado}

The \Corrado-spectrometer is situated in the Low Level Laboratory
at MPI-K at the shallow depth of 15~m~w.e. At this depth the
nucleonic component of the cosmic rays virtually vanishes and the
muon flux is reduced by a factor of $\sim$\nolinebreak3
\cite{heusser5}. The construction was completed in March 2007, and
testing and background measurements have been done. The detector's
shielding system is shown in figure \ref{corrado}.

\begin{figure}[h]
\centering
\includegraphics[width=0.35\textwidth]{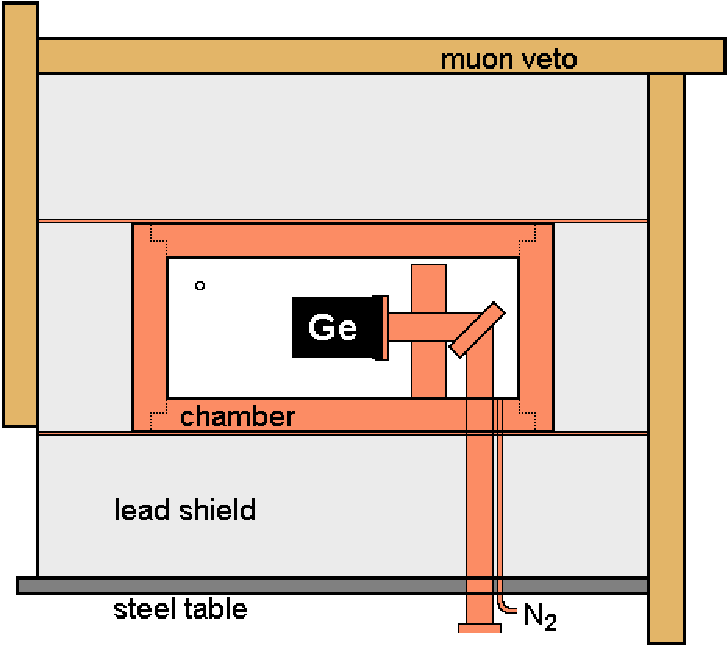}
\caption{Cross-section of the \Corrado-spectrometer.}
\label{corrado}
\end{figure}

The spectrometer is placed on a solid steel table with the
detector's dewar underneath. The HPGe crystal (coaxial, p-type,
mass 930~g\footnote{corresponding to a relative efficiency of
37\%}) is mounted in a cryostat made out of carefully selected
material. It has an aluminum end cap, and it's elbow shaped copper
neck is designed to avoid direct line of sight to external gamma
rays. In order to improve the Monte Carlo model of the cryostat
and crystal several x-ray images were taken before the insertion
into the shielding.

The sample chamber is made out of 5~cm thick electrolyte copper.
It allows for samples in Marinelli geometry with the dimensions
25$\times$25$\times$33~cm$^3$. The chamber can be evacuated and
flushed with nitrogen. During operation it is constantly flushed
by the boil-off nitrogen from a dewar to ensure a low and stable
$^{222}$Rn level in the chamber. Outside the chamber there is an
additional passive shield made of 15 to 20~cm low activity lead.
The innermost layer partly consits of lead from the Polish company
Plombum with a $^{210}$Pb contamination of 5~Bq/kg \cite{gempi2}.
A thicker shield is not reasonable since it would lead to a higher
neutron production due to cosmic rays \cite{heusser4}. All lead
bricks and copper parts of the shielding went through a thorough
etching procedure to remove surface contamination. The full
passive shield and flushing system supresses the background due to
environmental and airborne radioactivity by a factor
$\sim$\nolinebreak100 -- see figure \ref{UG_Vergleich}. The
shielding can be opened from the frontside to insert samples.

{\Corrado} is equipped with five multiwire proportional chambers
operated in anticoincidence with the Ge-diode as a muon veto. With
a dead time of 50~$\mu$s/signal it is set to reject background
from muon induced bremsstrahlung and short lived muon-activated
isotopes. A suppression factor of $\sim$\nolinebreak10 (88\%) has
been determined from the difference of the vetoed and the unvetoed
spectra (figure \ref{UG_Vergleich}) \cite{wernersdipl}. This is a
preliminary result and we seek to approach the high suppression
($>$93\%) of {\sc Dario} \cite{heusser1}. Alltogether a background
suppression of three orders of magnitude compared to the
unshielded detector in the LLL has been achieved.

\begin{table}[b]
\begin{minipage}[b]{7.5cm}
\centering
\begin{tabular}{c c c}
\hline Energy & Isotope & Counts per day \\ \hline 100-2700~keV &
& 4609~$\pm$~17~~ \\ ~$\,$609~keV & $^{214}$Bi & $<$1.9 \\
1333~keV & $^{~60}$Co & $<$2.1 \\ 1461~keV & $^{~40}$K~ &
~1.45~$\pm$~0.80
\\ 2615~keV & $^{208}$Tl & $<$1.3 \\ \hline
\end{tabular}
\caption{\Corrado-background countrates based on 16.7 days of
counting. Upper limits are given with 90\% confidence level.}
\label{bg corrado}
\end{minipage}
\hfill
\begin{minipage}[b]{7.5cm}
\centering
\begin{tabular}{c c c c c}
\hline \hspace{0.5cm} & Long lived & & Specific activity & \\ &
isotope & & [mBq/kg] & \hspace{0.5cm} \\ \hline & $^{226}$Ra & &
$<$2.38 & \\ & $^{228}$Ra & & $<$4.78 & \\ & $^{~40}$K~ & &
$<$24.99 & \\ & $^{228}$Th & & $<$1.14 & \\ \hline
\end{tabular}
\caption{Makrolon$^\circledR$ sample of 13~kg after 2.5 days
counting in \Corrado. Upper limits are given with 90\% confidence
level.} \label{makrolon}
\end{minipage}
\end{table}

The background spectrum with full shield reveals one background
line from $^{40}$K with $(1.45\pm0.80)$~cts/day. All other
isotopes lie below the detection limit -- results are given in
table \ref{bg corrado}. A long-term measurement with several weeks
of counting time is scheduled as soon as the ongoing renovation of
the LLL is completed. The aspired sensitivity of
$\sim$\nolinebreak1~mBq/kg has been achieved for $^{228}$Th with a
Makrolon$^\circledR$\footnote{a polycarbonate from Bayer
MaterialScience} sample of 13~kg and only 2.5 days of counting
time -- see table \ref{makrolon}.

\section{The Ge{\normalsize MPI} III-Spectrometer}

\subsection{The Design}

\begin{figure}[h]
\centering \includegraphics[width=0.4\textwidth]{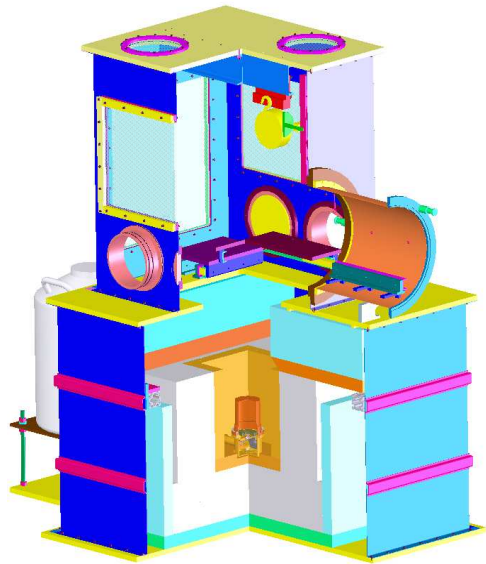}
\caption{Cutaway view of the \GempiIII~shielding and radon
protection system.} \label{GempiIII}
\end{figure}

{\GempiIII} is located in the LLRF at the LNGS with an overburden
of 3500~m~w.e. At this depth the muon flux is reduced by six
orders of magnitude to approximately 1~$\mu$/m$^2$h. This makes an
active muon shield dispensable. The assembly of the shielding was
completed in January 2007, but a subsequently discovered
$^{207}$Bi contamination postpones it's operation until today.
Figure \ref{GempiIII} shows a view of the spectrometer's shielding
system with the detector inside.

The cryostat is custom made from electro-refined copper and other
carefully selected and screened material. All materials were
chosen to minimize bulk contamination and to avoid high neutron
activation cross sections. It was stored underground between
fabrication and cleaning processes to reduce activation by cosmic
rays. The detector is a coaxial, p-type HPGe-crystal with a mass
of 2.3~kg. The conception and assembly of the cryostat system was
undertaken in close collaboration with Canberra Semiconductors
N.V., Olen, Belgium.

The sample chamber consists of 5~cm of NOSV copper \cite{nosv}
with the inner dimensions 25$\times$25$\times$30~cm$^3$. It is
surrounded by 20~cm of low activity lead. The innermost 5 to 10~cm
consist of lead from the Polish company Plombom (see section
\ref{corrado}). The shielding on top sits on rollers and may be
moved aside to have access to the chamber from above. Like for
{\Corrado} all lead and copper parts went through an etching
procedure to remove surface contamination. The shield is completed
by 5~cm thick slabs of borated (7\%) polyethylene from the bottom
and the four sides.

The setup is enclosed in a two-part air-tight stainless steel
casing for radon protection, which is constantly flushed at slight
overpressure with boil-off nitrogen. The bottom part envelopes the
shielding, whereas there is a sample storage and handling
compartment on top. It has acrylic windows, glove openings from
two sides, an airlock system for the insertion of samples, and a
movable table to transfer heavy loads. Therewith it is possible to
permanently operate the spectrometer without exposing the inner
volume to air. A space in the upper compartment is destined for
the storage of samples prior to their measurement, to allow for
the decay of $^{222}$Rn/$^{220}$Rn and their progenies. The entire
assembly of the \GempiIII-setup was done under cleanroom
conditions.

\subsection{Investigation on a $^{207}$Bi contamination}
\label{contamination}

\begin{figure}[h]
\begin{center}
\begin{minipage}[t]{7.7cm}
\includegraphics[width=0.7\textwidth, angle=-90]{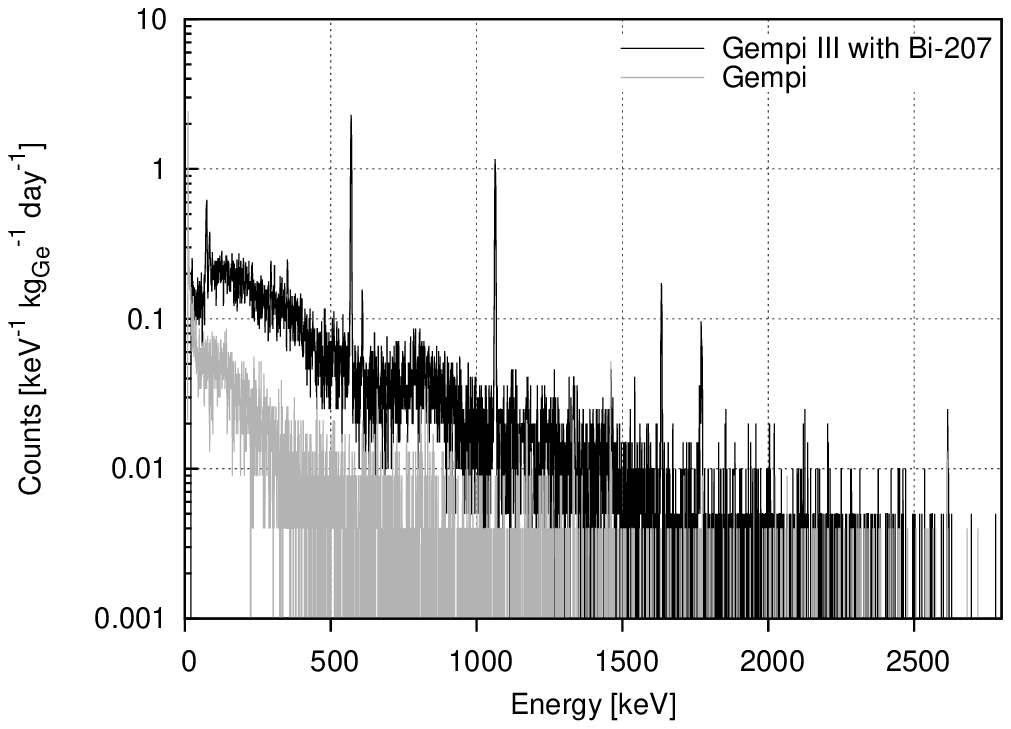}
\caption{Background spectrum of {\GempiIII} (black) with
$^{207}$Bi-contamination and provisional nitrogen flushing in
comparison to the Ge{\sc mpi}-spectrometer.} \label{Gempi_1u3}
\end{minipage}
\hfill
\begin{minipage}[t]{7.7cm}
\includegraphics[width=0.7\textwidth, angle=-90]{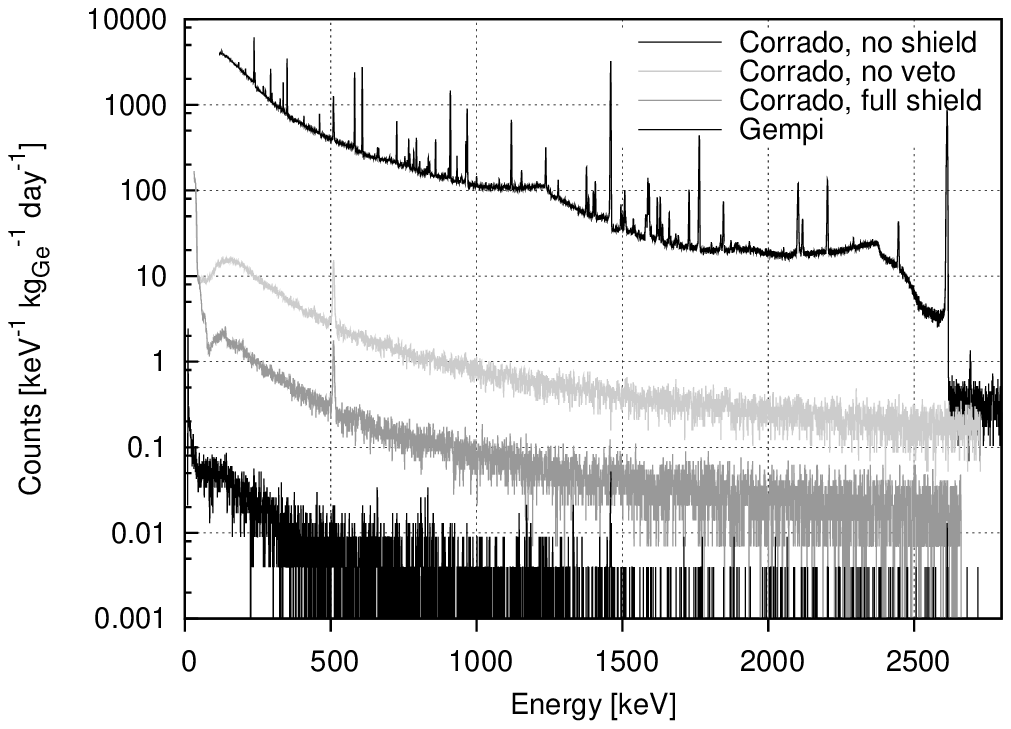}
\caption{Background spectra of {\Corrado} at MPI-K Heidelberg with
different shielding configurations in comparison to Ge{\sc mpi} at
Gran Sasso.} \label{UG_Vergleich}
\end{minipage}
\end{center}
\end{figure}

After comissioning of the spectrometer it turned out that it
suffers a $^{207}$Bi contamination. The two main lines are visible
with (37.8$\pm$1.6)~cts/d for 570~keV and (21.2$\pm$1.2)~cts/d for
1064~keV, as well as the summation peak with (3.0$\pm$0.5)~cts/d
-- see figure \ref{Gempi_1u3}. The occurrence of the latter
indicates a close proximity of the contamination to the crystal.
This is confirmed by measurements with additional shielding within
the sample chamber, leaving the denoted countrates unchanged. The
line ratio of the count rates for the 570~keV and 1064~keV line
allows to conclude that only minor attenuation takes place between
source and detector. The absence of conversion electrons in the
spectrum excludes a contamination inside the crystal's borehole.
Further investigation employing a summation peak analysis via MC
simulations to determine the distance towards the crystal is
pursued \cite{contam}. The activity was determined to be
$A(^{207}Bi)=(3.74\pm0.39)$~mBq.

$^{207}$Bi has no natural abundance, so possible candidates for
the contamination are cross-contamination with a source and a
$^{207}$Bi-production via (p,n)-reaction on $^{207}$Pb. The latter
was ruled out by the above analysis and a cross check by screening
the spare lead parts. After a long forensic struggle the origin of
the contamination was identfied as a cross contamination from a
hydrochloric acid solution containing $^{207}$Bi. It was
transferred via tweezers from an allegedly clean toolbox during
the cryostat-mounting, transfusing approximately 4~$\mu$g of
solution equivalent. The origin of the $^{207}$Bi-source lies
outside of Canberra's and our reference.

A similar $^{207}$Bi contamination has been discovered on the
Ge{\sc mpi IV}-cryostat in the course of this investigation.
However, it is lower by a factor of ten. A thorough etching
attempt for the cryostat and methanol baths for the crystal were
performed and tested at the end of July 2007, with the results
still to come.

\section{Conclusion}

The \Corrado-spectrometer shows, that a background supression by a
factor $\sim$\nolinebreak1000 is feasible in shallow depth. Apart
from a radiopure cryostat and passive shielding, an active muon
veto and radon protection of the sample chamber by constant
nitrogen flushing is needed. Further improvement of the muon veto
is intended. The spectrometer is designed for large samples and
has succeeded to meet the aspired sensitivity of
$\sim$\nolinebreak1~mBq/kg for $^{228}$Th. At this level it is
suitable for material screening within \Gerda.

The \GempiIII-spectrometer failed to reach the sensitivity of
Ge{\sc mpi}, which is the most sensitive $\gamma$-spectrometer
available for routine material screening. This is due to a
$^{207}$Bi contamination of (3.74$\pm$0.39)~mBq within the
cryostat of the detector. A variety of approaches are taken to
confine the possible locations, and a cleaning operation has been
carried out and tested on the Ge{\sc mpi IV}-detector.

\section*{Acknowledgements}
The technical staff of the MPI workshops is thanked for the
skillfull construction of the \GempiIII-cryostat and their
enduring commitment in realising the shielding systems of both
spectrometers. U. Schwan is thanked for her perpetual support and
counsel in surface etching and the contamination issue. We greatly
acknowledge the very good cooperation with Canberra Semiconductors
N.V., Olen, Belgium in the design and assembly of the
\GempiIII-cryostat.

This work was partly supported by INTAS (Project Nr.
05-1000008-7996) and by the DFG within the SFB Transregio 27
"Neutrinos and Beyond".

\end{document}